\documentclass[12pt]{revtex4}
\usepackage{amssymb}
\usepackage{epsf}
\usepackage[LCY]{fontenc}
\usepackage[cp866]{inputenc}
\usepackage[english]{babel}
\usepackage{rotating}
\newcommand{\df}{\,\mathrm{d}}
\begin{document}
\sloppy
\author{A.\,Bukin}
\email{bukin@inp.nsk.su}
\affiliation{Budker Institute of Nuclear Physics, Novosibirsk}
\date{\today}
\title{Fitting function for asymmetric peaks}
\begin{abstract}
In the paper a new fitting function is suggested, which
can essentially increase the existing instrumentation for
fitting of asymmetric peaks with the only maximum.
\end{abstract}

\maketitle

\section{Introduction}
During experimental data processing often there is a task to
fit arbitrary distribution with the only peak to some
function with a set of free parameters.
This can be necessary for more precise determination of peak position,
or full width at half maximum (devise resolution),
or for the signal form approximation using Monte Carlo events for subsequent
analysis of data distribution and so on.
Available set of fitting functions does not always matches the experimental
requirements, especially for high statistics (it is difficult to achieve
suitable confidence level).

In practice most often the sum of the three Gaussian distributions is
used to fit such histograms.
In most cases this fit provides suitable confidence level.
At least one can add more and more Gaussians until confidence level is
admissible.
Disadvantage of this method is that every Gaussian used is symmetric,
so in principle one can not used sum of Gaussians with common center
to fit asymmetric distributions, and if all Gaussians have different
centers, then it is difficult to provide the only maximum of fitting function
and monotony to the right and left of maximum.

Another often used function is a spline of two halves of different
Gaussians. This function is good in case that the function ``tails''
can be approximated with Gaussian.
However for high statistics the major experimental distributions far of peak
have exponential or power-of-x dependence, which do not match Gaussian.

Logarithmic Gaussian distribution is also often used (at SLAC it is called
``Novosibirsk'' function):
\begin{equation}
F_N(x;x_p,h,\lambda)=A\exp\left[
-\frac{1}{2\sigma^2}\ln^2\frac{x_m-x}{x_m-x_p}
\right],\;\; \int\limits_{\frac{x_m-x}{x_m-x_p}>0}\!\!\!\!\!F_N\df x=1,
\end{equation}
where $x_p$ is a peak location (function maximum),
 $h$ is full width at half maximum (FWHM),
  $\lambda$ is asymmetry parameter,
  $A$ is normalization factor.
\begin{equation}\label{eq:xm-Expression}
x_m=x_p-\frac{z+\frac{1}{z}}{2\lambda},
\end{equation}
\begin{equation}
\sigma=\frac{1}{\sqrt{2\ln 2}}\ln z,\;\;
A=\sqrt{\frac{2}{\pi}}\frac{\left|\lambda\right|}{\left(z+\frac{1}{z}\right)\sigma}
e^{-\frac{\sigma^2}{2}},
\end{equation}
\begin{equation}
z=\sqrt{h\left|\lambda\right|+\sqrt{h^2\lambda^2+1}}.
\end{equation}
As is seen from~(\ref{eq:xm-Expression}),
for $\lambda>0$ the boundary coordinate $x_m<x_p$ and all $x>x_m$.
If $\lambda<0$, then $x_m>x_p$ and $x<x_m$.
Using these notations one can rewrite this function, excluding
the inconvenient variable  $x_m$:
\begin{equation}
\begin{array}{l}
F_N(x;x_p,h,\lambda)=\sqrt{\frac{2}{\pi}}\cdot\frac{1}{z+\frac{1}{z}}\cdot
\frac{\left|\lambda\right|}{\sigma}
\exp\left[-\frac{\sigma^2}{2}
-\frac{1}{2\sigma^2}\ln^2\left(1+2z\lambda\frac{x-x_p}{z^2+1}\right)
\right].
\end{array}
\end{equation}
Function is equal to 0 for all  $x$,
for which the logarithm argument is negative.
Possible values of parameters:
$h>0$, $\lambda$ is arbitrary, $x_p$ is arbitrary. For $\lambda\sim 0$
the formulae have an ambiguity of the type $0/0$, so
some parameters should be expanded to the Tailor series.
\begin{equation}
z\begin{array}[t]{c}\sim \\[-2mm]\scriptstyle\lambda\sim 0\end{array}
1+\frac{h\left|\lambda\right|}{2}+\frac{h^2\lambda^2}{8},
\end{equation}
therefore
\begin{equation}
\frac{2\lambda z}{z^2+1}
\begin{array}[t]{c}\sim \\[-2mm]\scriptstyle\lambda\sim 0\end{array}
\lambda\cdot\left[1-\frac{h^2\lambda^2}{8}
+\frac{7h^4\lambda^4}{128}\right],
\end{equation}
\begin{equation}
\sigma^2
\begin{array}[t]{c}\sim \\[-2mm]\scriptstyle\lambda\sim 0\end{array}
\frac{h^2\lambda^2}{8\ln 2}\cdot\left[
1-\frac{h^2\lambda^2}{3}
\right],\;\;\; \frac{\sigma}{\left|\lambda\right|}
\begin{array}[t]{c}\sim \\[-2mm]\scriptstyle\lambda\sim 0\end{array}
\frac{h}{2\sqrt{2\ln 2}}\cdot\left[1-\frac{h^2\lambda^2}{6}
\right],
\end{equation}
\begin{equation}
\frac{\sigma^2}{2}
+\frac{1}{2\sigma^2}\ln^2\left(1+2z\lambda\frac{x-x_p}{z^2+1}\right)
\begin{array}[t]{c}\sim \\[-2mm]\scriptstyle\lambda\sim 0\end{array}
\frac{4\ln 2 \left(x-x_p\right)^2}{h^2}\cdot\left[
1-\left(x-x_p\right)\lambda
\right].
\end{equation}
For $\lambda=0$ the function converts to
\begin{equation}
F_N(x;x_p,h,0)=\sqrt{\frac{\ln 2}{\pi}}\cdot\frac{2}{h}\cdot
\exp\left[-\frac{4\ln 2 \left(x-x_p\right)^2}{h^2}
\right],
\end{equation}
that is Gaussian distribution with root mean square
$$
\sqrt{<(x-x_p)^2>}=\frac{h}{2\sqrt{2\ln 2}}\approx \frac{h}{2.3548}.
$$

This function is convenient for fitting the distributions
with abrupt spectrum end.
However many experimental distributions are more smooth,
and suggested in this paper function can be more successful.

\section{Convolution of Gaussian and exponential distributions}
It is suggested to build the fitting function on the base of
convolution of Gaussian and exponential distributions,
which can be easily derived:
\begin{equation}\label{eq:FitCurve1Explicit}
\begin{array}{l}
F_{B1}(x;x_g,\sigma_g,\lambda)=
\frac{1}{2\left|\lambda\right|}
\left[
1-\mathrm{erf}\left(-\frac{\left(x-x_g\right)\lambda}{\sigma_g\left|\lambda\right|\sqrt{2}}
+\frac{\sigma_g}{\left|\lambda\right|\sqrt{2}}
\right)
\right]
\times \\[3mm]\rule{35mm}{0mm}\times
\exp\left[
-\frac{x-x_g}{\lambda}
+\frac{\sigma_g^2}{2\lambda^2}
\right].
\end{array}
\end{equation}
Integral of this function over all $x$ equals 1.
Such a function was first used by the author in 2004 at SLAC
(BaBar Note \# 582)
for fitting the deposited energy distributions in calorimeter
with the aim of peak position and resolution determination
for the algorithms of absolute photon energy calibration,
and despite some technical difficulties, this function proved to be
enough convenient for fitting such distributions, especially for high statistics.

Technical difficulties appear when the argument of ERF function is big
\begin{equation}
\mathrm{erf}(x)=\frac{2}{\sqrt{\pi}}\int\limits_0^xe^{-\xi^2}\df\xi
\end{equation}
If the argument of ERF function in formula~(\ref{eq:FitCurve1Explicit})
is denoted as $z$
\begin{equation}
z=-\frac{\left(x-x_g\right)\lambda}{\sigma_g\left|\lambda\right|\sqrt{2}}
+\frac{\sigma_g}{\left|\lambda\right|\sqrt{2}},
\end{equation}
then the formula~(\ref{eq:FitCurve1Explicit})
looks like:
\begin{equation}\label{eq:FitCurve1Reduced}
F_{B1}(x;x_g,\sigma_g,\lambda)=
\frac{1}{2\left|\lambda\right|}
\left[
1-\mathrm{erf}\left(z
\right)
\right]
\exp\left[\frac{z\sigma_g\sqrt{2}}{\left|\lambda\right|}-
\frac{\sigma_g^2}{2\lambda^2}
\right].
\end{equation}
If $z\to +\infty$, then exponential index also goes to infinity,
and so the ambiguity of the type $0\cdot \infty$ arises.
Because of finite accuracy of computer calculations, this difficulty
appears rather fast, just for moderate values of $z\sim 10$.

\begin{figure}[tbp]
\epsfxsize=0.49\textwidth
\epsfbox{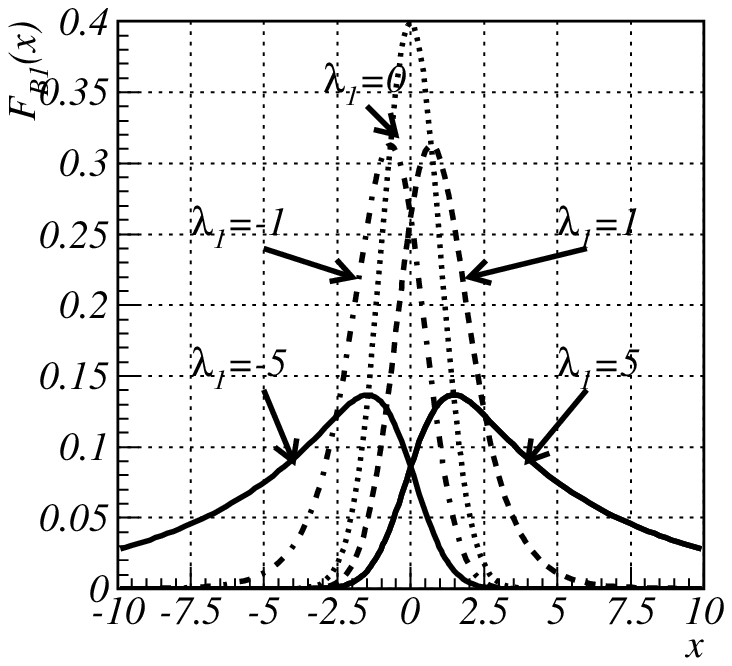}
\hfill
\epsfxsize=0.49\textwidth
\epsfbox{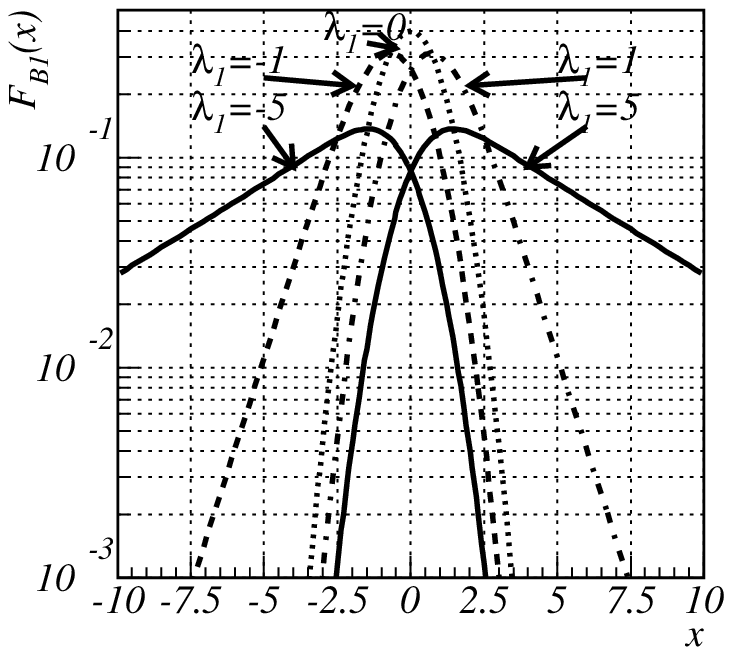}
\caption{\label{SampleOfFB1}Function $F_{B1}(x;x_g,\sigma_g,\lambda_1)$ plots for $\sigma_g=1,\;
x_g=0$ and several values of $\lambda_1$ }
\end{figure}

To avoid this problem one can use the asymptotic expansion
\begin{equation}
\begin{array}{l}
1-\mathrm{erf}\left(z\right)
\approx
\frac{e^{-z^2}}{z\sqrt{\pi}}\left[1+\sum\limits_{k=1}^{k_0}\frac{(-1)^k(2k-1)!!}{2^kz^{2k}}
\right]
,\;\;\; k_0\leq z^2.
\end{array}
\end{equation}
Substituting this expansion to~(\ref{eq:FitCurve1Reduced}), we obtain:
\begin{equation}\label{eq:FitCurve1Asymptot}
\begin{array}{l}
F_{B1}(x;x_g,\sigma_g,\lambda)=
\frac{1}{2\left|\lambda\right|z\sqrt{\pi}}
\exp\left[-\left(z-\frac{\sigma_g}{\sqrt{2}\left|\lambda\right|}\right)^2
\right]
\cdot
\left[1+\sum\limits_{k=1}^{k_0}\frac{(-1)^k(2k-1)!!}{2^kz^{2k}}
\right].
\end{array}
\end{equation}
Here for big  $z$ values no ambiguities appear,
function goes to 0.
This very expansion allows to find the limit for $
\lambda\to 0$.
Indeed,
\begin{equation}
\lim\limits_{\lambda\to 0} z=+\infty,\;\;\;
\lim\limits_{\lambda\to 0}
\left(z-\frac{\sigma_g}{\sqrt{2}\left|\lambda\right|}\right)=
\pm\frac{\left(x-x_g\right)}{\sigma_g\sqrt{2}},\;\;\;
\lim\limits_{\lambda\to 0} \left|\lambda\right|z=\frac{\sigma_g}{\sqrt{2}},
\end{equation}
and
\begin{equation}\label{eq:FitCurve1Lam=0}
F_{B1}(x;x_g,\sigma_g,0)=
\frac{1}{\sigma_g\sqrt{2\pi}}
\exp\left[-\frac{\left(x-x_g\right)^2}{2\sigma_g^2}
\right]
\end{equation}
Let us consider the limit $\sigma_g\to 0$.

\begin{equation}
\lim\limits_{\sigma_g\to 0} z=-\frac{x-x_g}{\lambda}\cdot \infty,
\;\;\; \lim\limits_{\sigma_g\to 0} z\sigma_g= -\frac{(x-x_g)|\lambda|}{\lambda\sqrt{2}},
\end{equation}
and substituting this to~(\ref{eq:FitCurve1Reduced}), we get
\begin{equation}\label{eq:FitCurve1Sigma=0}
F_{B1}(x;x_g,0,\lambda)=
\left\{
\begin{array}{l}
0,\;\;\frac{(x-x_g)}{\lambda}<0,\\[3mm]
\frac{1}{\left|\lambda\right|}\exp\left[-\frac{(x-x_g)}{\lambda}\right],
\;\; \frac{(x-x_g)}{\lambda}>0.
\end{array}\right.
\end{equation}
Function $F_{B1}$ plots for several sets of parameters
are presented in Fig.\,\ref{SampleOfFB1}.

Sometimes the integral distribution can be useful:
\begin{equation}
\begin{array}{l}
\Phi_{B1}(x;x_g,\sigma_g,\lambda)
=\!\!\!\int\limits_{-\infty}^x\!\!\!\df\xi\;F_{B1}(\xi;x_g,\sigma_g,\lambda)=
\frac{1}{2}\left[1+\mathrm{erf}\left(\frac{x-x_g}{\sqrt{2}\sigma_g}\right)\right]
-\\[1mm] \rule{40mm}{0mm}
-\frac{\lambda}{2\left|\lambda\right|}
e^{\frac{\sigma_g^2}{2\lambda^2}-\frac{x-x_g}{\lambda}}\left[
1-\mathrm{erf}\left(\frac{\sigma_g}{\sqrt{2}\left|\lambda\right|}-
\frac{\left(x-x_g\right)\left|\lambda\right|}{\sqrt{2}\sigma_g\lambda}
\right)\right].
\end{array}
\end{equation}

Usage of $F_{B1}$ to fit the distributions
would be more convenient, if the free parameter is $x_m$ ---
location of function maximum, instead of $x_g$.
Equation for the search for $x_m$ looks rather complicated:
\begin{equation}\label{eq:PeakCondition}\!\!\!\!
\begin{array}{l}
1-\mathrm{erf}\left(-\frac{\left(x_m-x_g\right)\lambda}{\sigma_g\left|\lambda\right|\sqrt{2}}
+\frac{\sigma_g}{\left|\lambda\right|\sqrt{2}}
\right)
=
\frac{\sqrt{2}\left|\lambda\right|}{\sigma_g\sqrt{\pi}}\exp\left[-
\left(-\frac{\left(x_m-x_g\right)\lambda}{\sigma_g\left|\lambda\right|\sqrt{2}}
+\frac{\sigma_g}{\left|\lambda\right|\sqrt{2}}
\right)^2
\right]
\end{array}
\end{equation}
or in other notations
\begin{equation}\label{eq:maxFb1}
1-\mathrm{erf}(z_m)=\rho\exp(-z_m^2),
\end{equation}
where
$\rho=\frac{\sqrt{2}\left|\lambda\right|}{\sigma_g\sqrt{\pi}}>0$.
For $\rho\to \infty$ $z_m\to -\infty$:
\begin{equation}
\begin{array}{l}
e^{z_m^2}=\frac{\rho}{2-\frac{e^{-z_m^2}}{|z_m|\sqrt{\pi}}\cdot\left[
1-\frac{1}{2z_m^2}\right]}
\Longrightarrow
z_m\approx -\sqrt{\ln\frac{\rho}{2-\frac{2}{\rho\sqrt{\left|\ln\frac{\rho}{2}\right|}
\sqrt{\pi}}\cdot\left[
1-\frac{1}{2\ln\frac{\rho}{2}}\right]}}
\Longrightarrow
\\[7mm]\rule{5mm}{0mm}\Longrightarrow
-\frac{\left(x_m-x_g\right)\lambda}{\sigma_g\left|\lambda\right|\sqrt{2}}
+\frac{\sigma_g}{\left|\lambda\right|\sqrt{2}}=
-\sqrt{\ln\frac{\left|\lambda\right|}{\sigma_g\sqrt{2\pi}}
}
\end{array}
\end{equation}
or
\begin{equation}
x_m-x_g\begin{array}[t]{c} \approx \\[-2mm] \scriptstyle \sigma_g \ll |\lambda|
\end{array}
\frac{\sigma_g^2}{\lambda}+\frac{\sigma_g\left|\lambda\right|\sqrt{2}}{\lambda}\cdot
\sqrt{\ln\frac{\left|\lambda\right|}{\sigma_g\sqrt{2\pi}}
}\to 0.
\end{equation}

For $\rho\ll 1$ $z_m\to +\infty$,
and here we also can derive approximate solution:
\begin{equation}
\frac{e^{-z_m^2}}{z_m\sqrt{\pi}}\left[1+\sum\limits_{k=1}^{k_0}\frac{(-1)^k(2k-1)!!}{2^kz_m^{2k}}
\right]
=\rho\exp\left(-z_m^2\right)
\end{equation}

\begin{equation}
\begin{array}{l}
z_m\approx \frac{1}{\rho\sqrt{\pi}}\cdot\left(1-\frac{\pi \rho^2}{2}\right)
\Longrightarrow
-\frac{\left(x_m-x_g\right)\lambda}{\sigma_g\left|\lambda\right|\sqrt{2}}
+\frac{\sigma_g}{\left|\lambda\right|\sqrt{2}}=
\frac{\sigma_g}{\sqrt{2}\left|\lambda\right|}
-\frac{\left|\lambda\right|}{\sqrt{2}\sigma_g}
\end{array}
\end{equation}
or
\begin{equation}
x_m-x_g\approx  \lambda\to 0.
\end{equation}
From~(\ref{eq:PeakCondition}) one can derive more terms of Taylor
series:
\begin{equation}
x_m-x_g\begin{array}[t]{c} \approx \\[-2mm] \scriptstyle |\lambda| \ll  \sigma_g
\end{array}
\lambda\cdot\left[1-\left(\frac{\lambda}{\sigma_g}\right)^2+
\left(\frac{\lambda}{\sigma_g}\right)^4+\ldots
\right].
\end{equation}

\begin{figure}[tbp]
\epsfxsize=0.7\textwidth
\centerline{\epsfbox{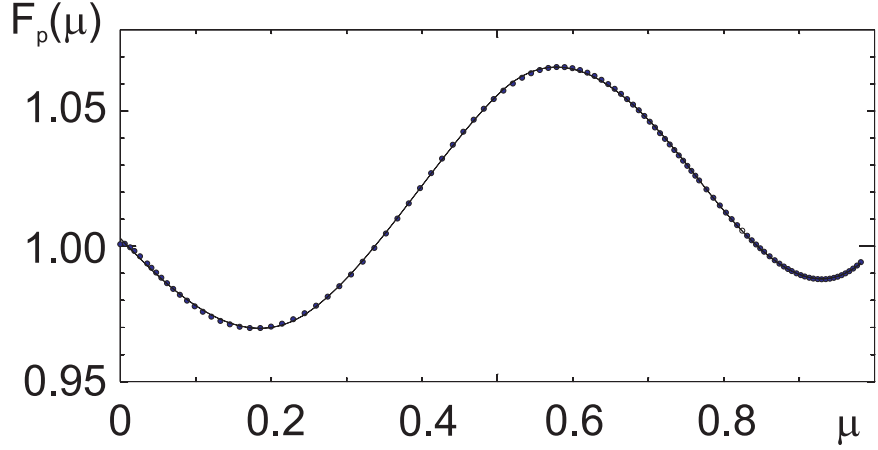}}
\caption{\label{FpApproximation}Function $F_p(\mu)$ plot (points)
and approximating cubic spline}
\end{figure}

Let us return to the equation~(\ref{eq:maxFb1}),
which should be solved in order to find  $F_{B1}$
maximum location. Let us transform the interval of
 $\rho$ variable to the interval  $(0,1)$:
\begin{equation}
\mu=e^{-\rho} \Longleftrightarrow \rho=-\ln \mu
\end{equation}
At the ends of the interval we know the solution:
\begin{equation}
\begin{array}{l}
\mu \sim 0 \Longrightarrow z_m\approx -\sqrt{\ln\left(-\frac{\ln\mu}{2}\right)}, \\[3mm]
\mu \sim 1\Longrightarrow z_m\approx \frac{\sqrt{\pi}\ln\mu}{2}-\frac{1}{\sqrt{\pi}\ln\mu}
\end{array}
\end{equation}
Let us look for approximating function in the form:
\begin{equation}
\begin{array}{l}
z_m=F_p(\mu)\cdot \frac{\left(1-\rho\right)
\left(8+\rho\right)\left(\mu+0.13\right)\left(1-\mu\right)}{\left(3+\rho\right)
\rho\sqrt{\pi}}
\sqrt{\frac{1}{\rho+1}+\pi\ln\left(1+\frac{\rho}
{2-\frac{2\ln\frac{\rho}{2}-1}{\left(\rho+2\right)\sqrt{\left|\ln\left(1+\frac{\rho}{2}\right)
\right|}^3
\sqrt{\pi}}}\right)}
\end{array}
\end{equation}
Function $F_p(\mu)$ plot and the approximating cubic
spline with 5 knots are presented in Fig.\,\ref{FpApproximation}.
Root mean square deviation equals $4.1\cdot 10^{-4}$,
maximum error of interpolation $2.2\cdot 10^{-3}$
is achieved at $\mu=1.0\cdot 10^{-5}$.
Spline coefficients are cited in Table\,\ref{tab:FpSpline}.
\begin{table}[tbp]
\caption{\label{tab:FpSpline}Spline of deficiency 2 coefficients for approximation
of $F_p(\mu)$ function.}
\begin{center}
\begin{tabular}{|l|c|c|c|c|c|}
\hline
Knot coordinate & 0 & 0.25 & 0.5 & 0.75 & 1 \\
\hline
Spline value    &1.00298 & 0.97574 & 1.05593 & 1.03058 & 0.99898 \\
\hline
Spline derivative    & -0.29472 & 0.19545 & 0.27768 & -0.35485 & 0.33206 \\
\hline
\end{tabular}
\end{center}
\end{table}
Now we can calculate the shift of peak position vs Gaussian center:
\begin{equation}
x_m-x_g=\frac{\sigma_g^2}{\lambda}-\frac{\sigma_g\left|\lambda\right|\sqrt{2}}
{\lambda}z_m=\Delta X_{mg}(\sigma_g,\lambda).
\end{equation}
The function $\Delta X_{mg}$
can be easily implemented in  any programming language,
using the above formulae.

\section{Build of fitting function on the base of $F_{B1}$}
In some cases the function  $F_{B1}$ itself can be suitable
for fitting.
However more often this function is not enough
flexible to provide satisfactory confidence level with
experimental distribution.

One could try to use the function which is the convolution of
the three distributions: Gaussian and two different exponential.
Such a function could be useful if the distribution ``tails''
both to the right and left from the peak do not match Gaussian
distribution.

However in this case and in other difficult ones the distributions
are fitted more successfully to the sum of different $F_{B1}$ functions
with the common parameter --- peak position $x_m$.
For the sum of two functions one can use the following expression:
\begin{equation}
\begin{array}{l}
F_{2B1}(x;x_m,\xi,\sigma_1,\lambda_1,\sigma_2,\lambda_2)=
\cos^2\xi\cdot F_{B1}\left(x;x_m-\Delta X_{mg}(\sigma_1,\lambda_1),\sigma_1,\lambda_1\right)+
\\[3mm]\rule{51mm}{0mm}+
\sin^2\xi\cdot  F_{B1}\left(x;x_m-\Delta X_{mg}(\sigma_2,\lambda_2),\sigma_2,\lambda_2\right)
\end{array}
\end{equation}

\begin{figure}[tbp]
\epsfxsize=0.49\textwidth
\epsfbox{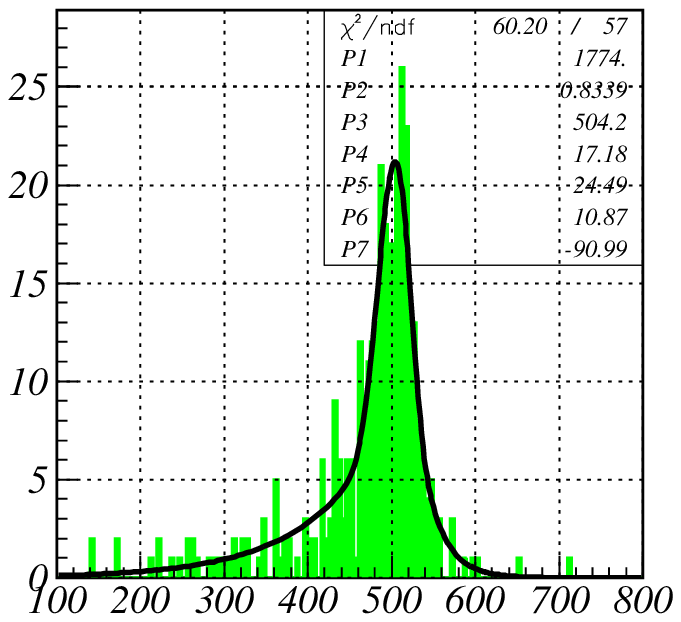}
\hfill
\epsfxsize=0.49\textwidth
\epsfbox{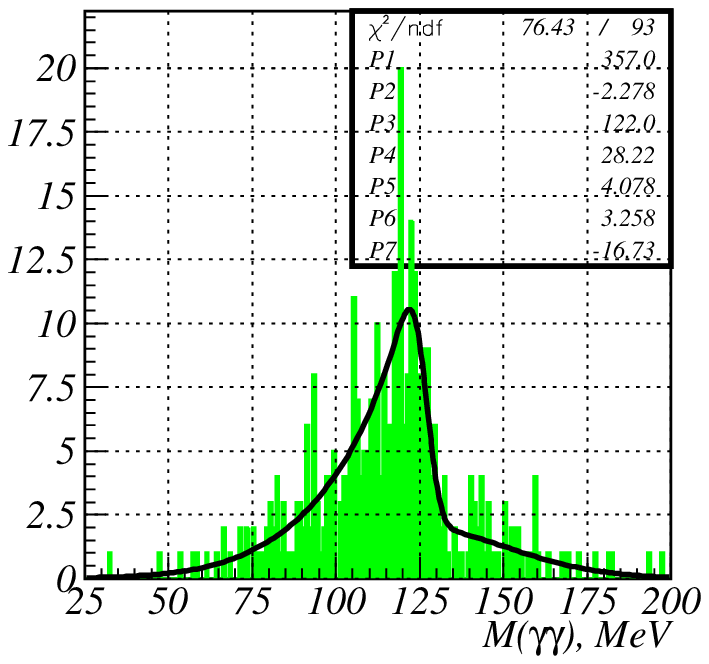}
\caption{\label{FitEtraM0acFB1}Distributions vs invariant masses
of photon pairs.
Left --- decay photons of $\eta$, right --- decay of $\pi^0$.
Fit function FIT2FB1.}
\end{figure}

Sample of using such a function is presented in Fig.\,\ref{FitEtraM0acFB1}
(one more parameter is added --- common factor $A$).

In principle for complicated cases one can use the sum of more $F_{B1}$
functions.
However for so many free parameters the likelihood function minimization
can be unstable, and one need to help MINUIT program.
The simplest and enough effective trick is
optimization of parameters in turn, initially fixed at some
reasonable values.

\section{Conclusion}
For fitting the smooth distributions with one peak
a function is suggested, which is the convolution of
Gaussian and exponential distributions.

In cases when this function is not enough flexible to provide
satisfactory confidence level, one can use
the sum of several functions with different parameters, but
the same peak position.

\end{document}